\documentclass[aps,prl,preprint,tightenlines,superscriptaddress,showpacs,byrevtex]{revtex4}
\usepackage{graphicx} 
\usepackage{dcolumn}  

\graphicspath{{ps}}


\newcommand{\ee}{e^{+}e^{-}}
\newcommand{\leplep}{\ell^{+}\ell^{-}}
\newcommand{\jp}{J/\psi}

\newcommand{\ecp}{\eta_{c}(2S)}
\newcommand{\psip}{\psi^{\prime}}
\newcommand{\onedtwo}{\eta_{c2}}
\newcommand{\threedtwo}{\psi_{2}(1{\rm D})}

\newcommand{\pipi}{\pi^{+}\pi^{-}}
\newcommand{\bbar}{B\bar{B}}
\newcommand{\kpi}{K^-\pi^{+}}
\newcommand{\ks}{K_{S}}

\newcommand{\Mbc}{M_{\rm bc}}
\newcommand{\DE}{\Delta E}

\newcommand{\rt}{\rightarrow}

\newcommand{\etal}{\em et al.}

\begin{document}

\preprint{\vbox{ \hbox{   }
                 \hbox{BELLE-CONF-0352}
}}

\title{ 
Observation of a new narrow charmonium state in exclusive $B^{\pm}\rt K^{\pm} \pipi\jp$ decays
}

\affiliation{Aomori University, Aomori}
\affiliation{Budker Institute of Nuclear Physics, Novosibirsk}
\affiliation{Chiba University, Chiba}
\affiliation{Chuo University, Tokyo}
\affiliation{University of Cincinnati, Cincinnati, Ohio 45221}
\affiliation{University of Frankfurt, Frankfurt}
\affiliation{Gyeongsang National University, Chinju}
\affiliation{University of Hawaii, Honolulu, Hawaii 96822}
\affiliation{High Energy Accelerator Research Organization (KEK), Tsukuba}
\affiliation{Hiroshima Institute of Technology, Hiroshima}
\affiliation{Institute of High Energy Physics, Chinese Academy of Sciences, Beijing}
\affiliation{Institute of High Energy Physics, Vienna}
\affiliation{Institute for Theoretical and Experimental Physics, Moscow}
\affiliation{J. Stefan Institute, Ljubljana}
\affiliation{Kanagawa University, Yokohama}
\affiliation{Korea University, Seoul}
\affiliation{Kyoto University, Kyoto}
\affiliation{Kyungpook National University, Taegu}
\affiliation{Institut de Physique des Hautes \'Energies, Universit\'e de Lausanne, Lausanne}
\affiliation{University of Ljubljana, Ljubljana}
\affiliation{University of Maribor, Maribor}
\affiliation{University of Melbourne, Victoria}
\affiliation{Nagoya University, Nagoya}
\affiliation{Nara Women's University, Nara}
\affiliation{National Kaohsiung Normal University, Kaohsiung}
\affiliation{National Lien-Ho Institute of Technology, Miao Li}
\affiliation{Department of Physics, National Taiwan University, Taipei}
\affiliation{H. Niewodniczanski Institute of Nuclear Physics, Krakow}
\affiliation{Nihon Dental College, Niigata}
\affiliation{Niigata University, Niigata}
\affiliation{Osaka City University, Osaka}
\affiliation{Osaka University, Osaka}
\affiliation{Panjab University, Chandigarh}
\affiliation{Peking University, Beijing}
\affiliation{Princeton University, Princeton, New Jersey 08545}
\affiliation{RIKEN BNL Research Center, Upton, New York 11973}
\affiliation{Saga University, Saga}
\affiliation{University of Science and Technology of China, Hefei}
\affiliation{Seoul National University, Seoul}
\affiliation{Sungkyunkwan University, Suwon}
\affiliation{University of Sydney, Sydney NSW}
\affiliation{Tata Institute of Fundamental Research, Bombay}
\affiliation{Toho University, Funabashi}
\affiliation{Tohoku Gakuin University, Tagajo}
\affiliation{Tohoku University, Sendai}
\affiliation{Department of Physics, University of Tokyo, Tokyo}
\affiliation{Tokyo Institute of Technology, Tokyo}
\affiliation{Tokyo Metropolitan University, Tokyo}
\affiliation{Tokyo University of Agriculture and Technology, Tokyo}
\affiliation{Toyama National College of Maritime Technology, Toyama}
\affiliation{University of Tsukuba, Tsukuba}
\affiliation{Utkal University, Bhubaneswer}
\affiliation{Virginia Polytechnic Institute and State University, Blacksburg, Virginia 24061}
\affiliation{Yokkaichi University, Yokkaichi}
\affiliation{Yonsei University, Seoul}
  \author{K.~Abe}\affiliation{High Energy Accelerator Research Organization (KEK), Tsukuba} 
  \author{K.~Abe}\affiliation{Tohoku Gakuin University, Tagajo} 
  \author{N.~Abe}\affiliation{Tokyo Institute of Technology, Tokyo} 
  \author{R.~Abe}\affiliation{Niigata University, Niigata} 
  \author{T.~Abe}\affiliation{High Energy Accelerator Research Organization (KEK), Tsukuba} 
  \author{I.~Adachi}\affiliation{High Energy Accelerator Research Organization (KEK), Tsukuba} 
  \author{Byoung~Sup~Ahn}\affiliation{Korea University, Seoul} 
  \author{H.~Aihara}\affiliation{Department of Physics, University of Tokyo, Tokyo} 
  \author{M.~Akatsu}\affiliation{Nagoya University, Nagoya} 
  \author{M.~Asai}\affiliation{Hiroshima Institute of Technology, Hiroshima} 
% ...
\collaboration{The Belle Collaboration}

\noaffiliation

\begin{abstract}

We report the first observation of a narrow charmonium state produced
in the exclusive decay process $B^{\pm}\rt K^{\pm}\pipi\jp$.  This state, which decays
into $\pipi\jp$, has a mass of $3871.8\pm 0.7 {\rm (stat)} \pm 0.4{\rm (syst)}$~MeV,
which is very near the $M_{D} + M_{D^*}$ mass threshold.
The results are based on an analysis of 152M $\bbar$ events
collected at the $\Upsilon(4S)$ resonance in the Belle detector at the 
KEKB collider.

\end{abstract}

\maketitle
\tighten

\section{Introduction}

A major experimental issue for the
$c\overline{c}$  charmonium particle
system is the missing (i.e., still unobserved) charmonium states that are expected to be
below threshold for decays to open charm and, thus, are narrow.
These include the $n=1$ singlet~P state, the $h_c(1{\rm P}),$ and,
possibly, the $J^{PC}=2^{-+}~1^1{\rm D}_2~\onedtwo$ and
$J^{PC}=2^{--}~1^3{\rm D}_2~\threedtwo$, the $n=1$ singlet and triplet
spin-2~D states, which are narrow if their masses are below
the $D\bar{D^*}$ threshold.  The
observation of these states and the determination of
their masses would provide useful information about
the spin-dependence of the charmonium potential.

$B$ meson decays are a prolific source of charmonium particles.
The large $B$ meson samples produced at $B$-factories provide
excellent opportunities to search for the missing states.
The Belle group recently reported the first observation of the
$\ecp$ via its $\ks\kpi$ decay channel in exclusive $B\rt K\ks\kpi$
decays based on an analysis of 44.8M~$\bbar$ events~\cite{skchoi}.
Strategies for finding the remaining missing states have been
presented by Eichten, Lane and Quigg~\cite{eichten}; they note that
a narrow $\threedtwo$ should have a substantial decay branching fraction
for the $\pipi\jp$ final state.  In this paper, we report on a
search for a narrow peak in the $\pipi\jp$ mass spectrum from exclusive
$B^+\rt K^+\pipi\jp$ decays~\cite{chargeconjugate}
using a 152M~$\bbar$ event sample.  The data
were collected in the Belle detector at the KEKB
energy-asymmetric $\ee$ collider, which operates at the $\Upsilon(4S)$
resonance ($\sqrt{s}=10.58$~GeV). KEKB is described in detail in 
ref.~\cite{kurokawa}; Belle is described in detail
in ref.~\cite{abashian}.

\section{$B$-meson signal selection}

We use events  that satisfy
the Belle-standard hadronic-event selection requirements
and have a pair of well identified oppositely charged
electrons or muons with an invariant mass in the 
range $3.076<M_{\leplep}<3.116$~GeV, a loosely identified
charged kaon and a pair of oppositely charged pions.
In order to reject background from $\gamma$ conversion
products and curling tracks, we require the $\pipi$ invariant
mass to be greater than 0.4~GeV.
To reduce the level of $\ee\rt q\bar{q}$ ($q=u,d,s~{\rm or}~c$-quarks)
continuum events in the sample,
we also require  $R_2 < 0.4$, where $R_2$ is the normalized
Fox-Wolfram moment~\cite{fox}, and $|\cos\theta_B|<0.8$, where
$\theta_B$ is the polar angle of the $B$-meson direction
in the center-of-mass system (cms).
 
Candidate $B^+\rt K^+\pipi\jp$
mesons are reconstructed using the energy difference
$\DE\equiv E_B^{\rm cms} - E_{\rm beam}^{\rm cms}$
and the beam-energy constrained
mass $\Mbc\equiv\sqrt{(E_{\rm beam}^{\rm cms})^2-(p_B^{\rm
cms})^2}$,   
where $E_{\rm beam}^{\rm cms}$ is the beam
energy in the cms,
and $E_B^{\rm cms}$ and $p_B^{\rm cms}$ are the cms energy and
momentum of the $B$ candidate.
The signal region is defined as 5.271 $< \Mbc <$ 5.289 GeV/$c^2$
and $|\DE |< $ 0.030 GeV, which corresponds to $\pm 3\sigma$
and $\pm 2.5\sigma$ from the central values, respectively.

\begin{figure}[htb]
\includegraphics[width=0.8\textwidth, height=0.4\textwidth]{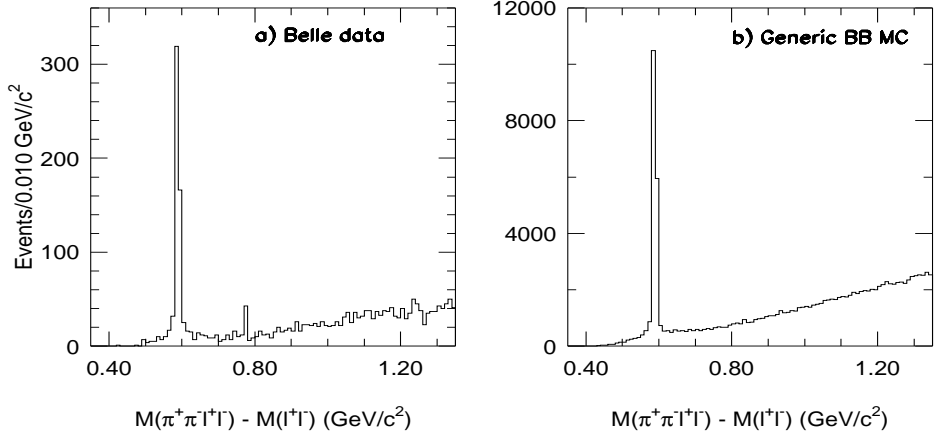}
\vspace{-0.8cm}
\caption{Distribution of $M(\pipi\leplep)-M(\leplep)$ for events
in the $\DE$-$\Mbc$ signal box for {\bf (a)} Belle data and {\bf (b)}
generic $\bbar$ MC events.}
\label{fig:pipill-ll} 
\end{figure}

\section{Results}

Figure~\ref{fig:pipill-ll}(a) shows the $\Delta M \equiv M(\pipi\leplep) - M(\leplep)$
mass distribution for events in the $\DE$-$\Mbc$ signal region.
Here a large peak corresponding to $\psip\rt\pipi\jp$ is evident near
0.59~GeV.  In addition there is a significant spike in the distribution near
0.77~GeV.  Figure~\ref{fig:pipill-ll}(b) shows the $\Delta M$
distribution for a large sample of generic $\bbar$ Monte Carlo (MC) events.  Except for
the prominent $\psip$ peak, the distribution is smooth and featureless.
In the rest of this paper we use $M(\pipi\jp)$ determined from $\Delta M + M_{\jp}$,
where $M_{\jp}$ is PDG~\cite{PDG} value for the $\jp$ mass.

\begin{figure}[htb]
\includegraphics[width=0.65\textwidth, height=0.53\textwidth]{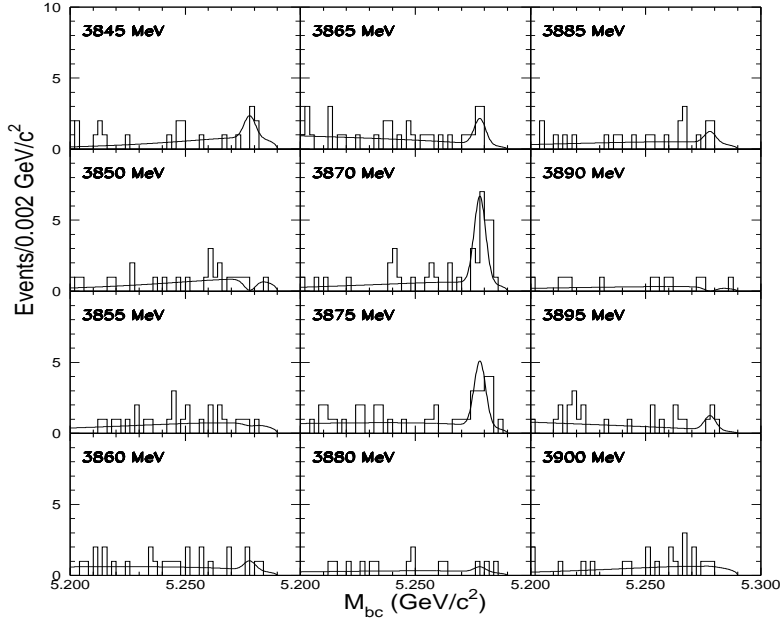}
\vspace{-0.8cm}
\caption{ $\Mbc$ distributions for events
in the $\DE$ signal box for 5-MeV $M(\pipi\jp)$ mass bins 
in the vicinity of the $\Delta M=0.77$~GeV
spike. The mass values listed in each panel correspond to
the center of the bin. The
curves are the results of fits described in the text.}
\label{fig:12box}
\end{figure}

In Fig.~\ref{fig:12box}, we show the $\Mbc$ distributions for 5-MeV $M(\pipi\jp)$ bins
for the $\Delta M$ region around the spike at $0.77$~GeV.  Here the $|\DE |<0.03$~GeV requirement
has been applied.  Significant $B$-meson signals
appear in the  $M(\pipi\jp)= 3870$~MeV mass bin and its immediately
adjacent bins, but nowhere else.  The curves
in the figures indicate the results of fits to the distributions using a Gaussian
function to represent the signal and the ARGUS function for the background~\cite{ARGUS}.  Here
the peak and width of the signal Gaussian are fixed at values determined from
fitting $\Mbc$ distributions in the $\psip$ peak region.

\subsection{Fit results}

\begin{figure}[htb]
\includegraphics[width=0.9\textwidth, height=0.4\textwidth]{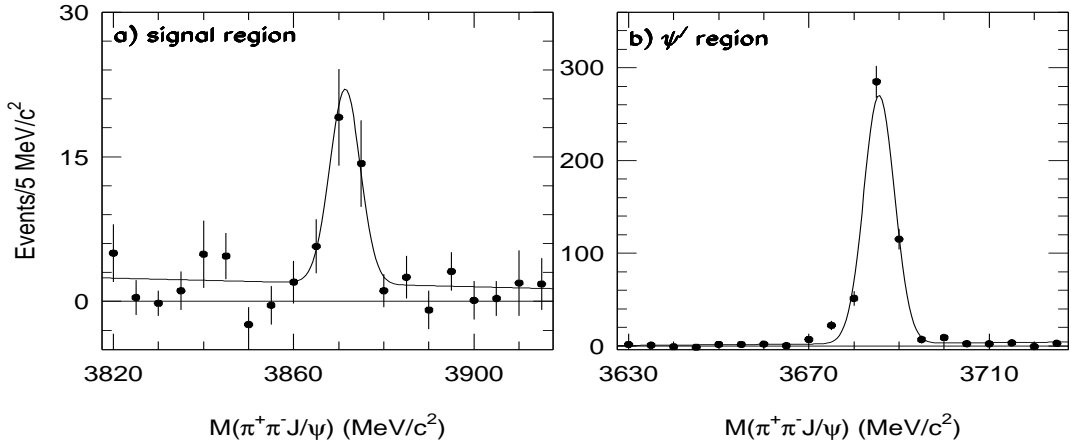}
\vspace{-0.8cm}
\caption{ The data points show the $B$-meson signal yields
from fits to the individual $\Mbc$ distributions in the
{\bf (a)} $M(\pipi\jp)=3870$~MeV and {\bf (b)} $\psip$
regions.  The curve represents the results of the fit described in the text.
}
\label{fig:slice_fit}    
\end{figure}

The $B$-meson signal yields from the fits to the individual $\Mbc$ distributions are shown 
as the data points in Fig.~\ref{fig:slice_fit}(a).   The curve is the result of a fit
with a Gaussian signal function and a second-order polynomial to
represent the background.  The fit gives a signal yield of $34.4\pm 6.5$~events, 
a peak mass of $M^{meas}_X = 3871.4 \pm 0.7$~MeV, 
and a width of $\sigma = 3.4\pm 0.5$~MeV.
The signal significance is estimated 
to be $\sqrt{-2\ln({\mathcal L}_0/{\mathcal L}_{max})} = 8.6\sigma$,
where ${\mathcal L}_{max}$ and ${\mathcal L}_0$ are the likelihood values for
the fitted signal yield and for  a fit with the 
amplitude of the signal function fixed at zero, respectively.
For comparison, we show in Fig.~\ref{fig:slice_fit}(b) the corresponding
distribution and fit for the $\psip$ peak region.  Here the fit results are:
$456.6\pm 22.4$ events; $M^{meas}_{\psip} = 3685.6 \pm 0.2$~MeV; and $\sigma=3.4\pm0.2$~MeV.
The fitted widths for the $\psip$ and signal peaks are in good agreement;
the fitted $\psip$ mass is 0.4~MeV below the PDG value of $M^{PDG}_{\psip} = 3686.00\pm 0.09$~MeV.

\subsection{Mass determination and limits on the total width}

We determine the mass of the signal peak by referencing to the well measured
$\psip $ mass:  
$$
M_X = M^{meas}_X - M^{meas}_{\psip} + M^{PDG}_{\psip} = 3871.8\pm0.7~{\rm MeV}.
$$
Here the error is statistical only.
Since we reference $M_X$ to the precisely known PDG value for the $\psip$ mass,
the systematic error is small.  The $M_{\psip}$ measurement, which
references the $\jp$ mass that is 590~MeV away,  is within  0.4~MeV
of the nominal PDG value.  Variation of the mass scale from
$M_{\psip}$ to the $M_{X}$ requires an extrapolation of only
186~MeV and, thus, can safely be expected to be less than this 
amount. We conservatively assign 0.4~MeV as the systematic error on the mass.

The fitted signal width is consistent with the resolution.  To determine a
90\% confidence-level (CL) upper
limit on the total width, we repeat fits to the distribution in
Fig.~\ref{fig:slice_fit}(a) using a resolution-broadened Breit-Wigner (BW)
function to represent the signal.  From the variation in $-2\ln({\mathcal L})$
with the input BW width we determine a 90\% CL upper limit of $\Gamma < 3.5$~MeV.

\subsection{Di-pion mass distribution and product branching fraction}

\begin{figure}[htb]
\includegraphics[width=0.9\textwidth, height=0.4\textwidth]{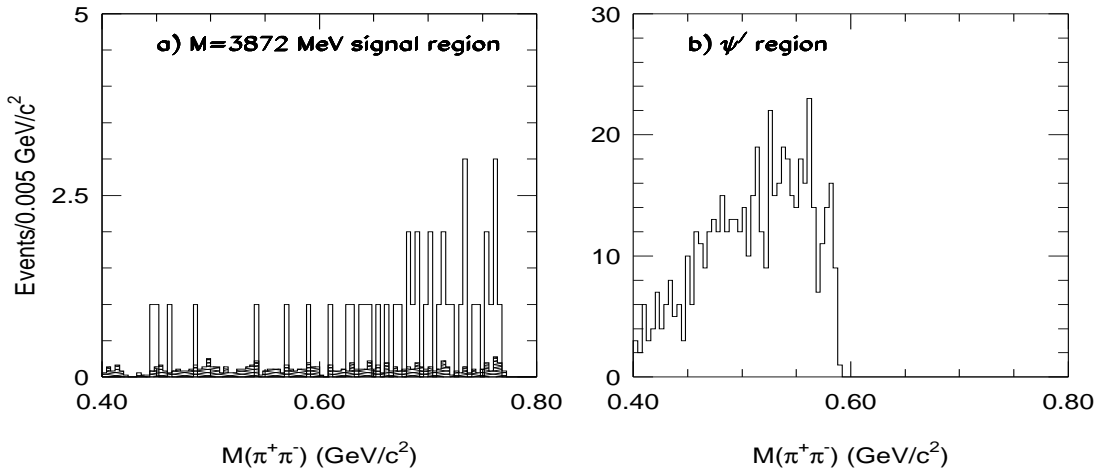}
\vspace{-0.8cm}
\caption{  The $M(\pipi)$  distributions for events in the
{\bf (a)} $M(\pipi\jp)=3870$~MeV signal region and {\bf (b)}
the $\psip$ region.  The shaded histogram is the sideband data
normalized to the signal-box area.}
\label{fig:mpipi}    
\end{figure}

The open histogram in Fig.~\ref{fig:mpipi}(a) shows the $\pipi$ invariant mass distribution
for the $M=3872$~MeV signal region;  the shaded histogram shows the  corresponding
distribution for events in the non-signal $\DE$-$\Mbc$ region, normalized 
to the signal area.   The $\pipi$ invariant masses tend to
cluster near the kinematic boundary, which is around the $\rho$ mass; the
entries below the $\rho$ are consistent with background.  Figure~\ref{fig:mpipi}(b)
shows the $\pipi$ mass distribution for the $\psip$ events.  This distribution 
also peaks near the upper kinematic boundary, which in this case is near $590$~MeV.

Assuming equal acceptances for both channels,
we determine a ratio of product branching fractions for
$B^+\rt K^+ X$, $X\rt \pipi\jp$ and $B^+\rt K^+ \psip$, $\psip \rt \pipi\jp$ 
to be
$$
\frac{{\cal B}(B^+\rt K^+ X)\times{\cal B}(X\rt\pipi\jp)}
{{\cal B}(B^+\rt K^+ \psip)\times{\cal B}(\psip\rt\pipi\jp)}   
= 0.075 \pm 0.014{\rm (stat)} \pm 0.007{\rm (syst)}.
$$
Here the systematic error includes possible different
efficiencies for the $M(\pipi)>0.4$~GeV selection
requirement for the two channels, estimated by
comparing the $\psip$ yield with and without this
requirement.

\section{Discussion}

The mass of the observed state is higher than potential model
expectations for the center-of-gravity ({\it cog}) of the $1^3{\rm D}_{cJ}$ states:
the Cornell~\cite{cornell} and the Buchm\"uller-Tye~\cite{tye} potentials
both give $M_{\rm cog}(1{\rm D})=3810$~MeV, which is 60~MeV below
the observed state. (As the center state of the $^3{\rm D}_{cJ}$ triplet,
one expects the $^3{\rm D}_{c2}$ to be near the {\it cog}.)  Moreover, the
$^3{\rm D}_{c1}$ state has been identified with the $\psi^{\prime \prime}(3770)$.
Identification of the observed state with the $^3{\rm D}_{c2}$ state would
imply a $^3{\rm D}_{c2}-^3{\rm D}_{c1}$ splitting of $\sim 100$~MeV.   On the
other hand, complications due to  effects of coupling to
real $D\bar{D}$ and virtual $D\bar{D^*}$ states may reduce the
validity of potential model calculations for these states~\cite{cornell,rosner,quigg}.

The $\pipi$ invariant masses for the $M=3872$~MeV signal region concentrate near
the upper kinematic boundary (Fig.~\ref{fig:mpipi}(a)), as is also the case for 
$\psi'\rt\pipi\jp$ (Fig.~\ref{fig:mpipi}(b)).  For the $M=3872$~MeV signal region, however, 
the boundary corresponds to the mass of the $\rho$.  The decay $^3D_{c2}\rt \rho~\jp$ is
an isospin-violating process, and these are strongly suppressed in
$\psip\rt\jp$ transitions.  With more data, we could determine whether or
not the $\pipi$ system is coming from $\rho$ meson decay, which could have important
consequences for the interpretation of this state.

The measured mass of the state is within errors of the $D^0\bar{D^{*0}}$ mass threshold
($3871.3 \pm 0.5$~MeV~\cite{PDG}). This would be
expected for a loosely bound $D\bar{D^*}$ multiquark ``molecular state,'' such as proposed 
by De Rujula, Georgi and  Glashow in 1977~\cite{georgi}.

\section{Conclusion}

We have observed a strong signal ($8.6\sigma$) for a state that
decays to $\pipi\jp$ with 
\begin{eqnarray*}
M                & = &    3871.8 \pm 0.7~{\rm (stat)} \pm 0.4~{\rm (syst)}~{\rm MeV}\\   
\Gamma           & < &    3.5~{\rm MeV}~.
\end{eqnarray*}
\noindent
This mass value is about 60~MeV higher than potential model predictions
for a 1D charmonium state  and equal, within errors,  to $M_{D^0} + M_{D^{*0}}$.
This coincidence with the $D^0 \bar{D^{*0}}$ mass threshold  suggests that this
may be  a $DD^*$ multiquark state.

\section{Acknowledgements}

\noindent
We wish to thank the KEKB accelerator group for the excellent
operation of the KEKB accelerator.
We acknowledge support from the Ministry of Education,
Culture, Sports, Science, and Technology of Japan
and the Japan Society for the Promotion of Science;
the Australian Research Council
and the Australian Department of Industry, Science and Resources;
the National Science Foundation of China under contract No.~10175071;
the Department of Science and Technology of India;
the BK21 program of the Ministry of Education of Korea
and the CHEP SRC program of the Korea Science and Engineering
Foundation;
the Polish State Committee for Scientific Research
under contract No.~2P03B 01324;
the Ministry of Science and Technology of the Russian Federation;
the Ministry of Education, Science and Sport of the Republic of
Slovenia;
the National Science Council and the Ministry of Education of Taiwan;
and the U.S.\ Department of Energy.



\begin{thebibliography}{99}
\bibitem{chargeconjugate} In this paper, the inclusion
of charge conjugate states is always implied.
\end{thebibliography}
\end{document}